**An Uncertainty Aided Framework for Learning based Liver $T_{1\rho}$ Mapping and Analysis**

Chaoxing Huang[1,2], Vincent Wai-Sun Wong[3], Queenie Chan[4], Winnie Chiu-Wing Chu[1,2], Weitian Chen[1,2]

1. Department of Imaging and Interventional Radiology, The Chinese University of Hong Kong, Hong Kong SAR
2. CUHK Lab of AI in Radiology (CLAIR), Hong Kong SAR
3. Department of Medicine & Therapeutics, The Chinese University of Hong Kong, Hong Kong SAR
4. Philips Healthcare, Hong Kong SAR

Corresponding author:

Weitian Chen

Room 15, Sir Yue Kong Pao Centre for Cancer

Prince of Wales Hospital Shatin, NT

Hong Kong

(852)-3505-1036

Email: wtchen@cuhk.edu.hk

Keywords: Quantitative MRI, Deep Learning, Uncertainty, $T_{1\rho}$

**Abstract**

**Objective**: Quantitative $T_{1\rho}$ imaging has potential for assessment of biochemical alterations of liver pathologies. Deep learning methods have been employed to accelerate quantitative $T_{1\rho}$ imaging. To employ artificial intelligence-based quantitative imaging methods in complicated clinical environment, it is valuable to estimate the uncertainty of the predicated $T_{1\rho}$ values to provide the confidence level of the quantification results. The uncertainty should also be utilized to aid the post-hoc quantitative analysis and model learning tasks. **Approach**: To address this need, we propose a parametric map refinement approach for learning-based $T_{1\rho}$ mapping and train the model in a probabilistic way to model the uncertainty. We also propose to utilize the uncertainty map to spatially weight the training of an improved $T_{1\rho}$ mapping network to further improve the mapping performance and to remove pixels with unreliable $T_{1\rho}$ values in the region of interest. The framework was tested on a dataset of 51 patients with different liver fibrosis stages. **Main results**: Our results indicate that the learning-based map refinement method leads to a relative mapping error of less than 3% and provides uncertainty estimation simultaneously. The estimated uncertainty reflects the actual error level, and it can be used to further reduce relative $T_{1\rho}$ mapping error to 2.60% as well as removing unreliable pixels in the region of interest effectively. **Significance**: Our studies demonstrate the proposed approach has potential to provide a learning-based quantitative MRI system for trustworthy $T_{1\rho}$ mapping of the liver.

## 1. INTRODUCTION

Quantitative MRI (qMRI) methods are important non-invasive technologies for assessment of diseases(Gulani and Seiberlich, 2020, Cristinacce et al., 2022). Quantitative $T_{1\rho}$ imaging is a promising qMRI technique with applications in various diseases. It has emerged as a potential biomarker for evaluating liver inflammation(Arihara et al., 2022, Takayama et al., 2022). Dynamic glucose enhanced $T_{1\rho}$ imaging also shows potential in mapping glucose metabolism in the liver(Qian et al., 2023). However, quantitative $T_{1\rho}$ imaging requires acquisition of multiple contrast images at different times of spin-lock (TSL), resulting in prolonged scan time and reduced temporal resolution compared to anatomical imaging. Various deep learning-based methods have been proposed to accelerate quantitative $T_{1\rho}$ imaging either by under-sampling k-space data or reducing the number of $T_{1\rho}$-weighted images used for mapping(Li et al., 2023, Huang et al., 2022a, Liu et al., 2022, Huang et al., 2022b). It is desirable to use as few $T_{1\rho}$-weighted images as possible to perform $T_{1\rho}$ mapping to optimize scan efficiency. However, the existing learning-based $T_{1\rho}$ mapping methods developed for this purpose in liver imaging still have relatively large errors(Huang et al., 2022a, Huang et al., 2022b). Further work is needed to improve the performance of deep learning-based $T_{1\rho}$ mapping in the liver to make it available for routine clinical use.

To deploy learning-based $T_{1\rho}$ mapping systems in a clinical environment, it is important to provide a confidence level or uncertainty estimation of the prediction results to aid human-machine interaction and decision-making. Most current learning-based qMRI mapping methods provide measurements in a deterministic manner without uncertainty estimations(Feng et al., 2022). Recently, some qMRI mapping algorithms have integrated uncertainty into the learning process, enabling the network to provide both the quantification results and the associated uncertainty level simultaneously. Huang et al.(Huang et al., 2022b) used self-supervised learning in the image domain to obtain a $T_{1\rho}$ map of the liver and its uncertainty estimations. However, the uncertainty values estimated in this work do not provide a direct estimation of the confidence of $T_{1\rho}$ quantification, as the uncertainty estimation is in the signal domain and its unit is not consistent with the relaxation time. Shih et al.(Shih et al., 2023) developed an uncertainty-aware supervised network to predict the uncertainty of proton-density fat fraction and R2*. Similarly, uncertainty was quantified for MRI parameter mapping of CEST imaging under 3T and 7T(Hunger et al., 2023, Glang et al., 2020).

Uncertainty maps of measured tissue parameters also contain valuable information, which can be used to improve the reliability and robustness of qMRI in clinical settings. However, there are few studies which discuss the utility of uncertainty maps in qMRI. In contrast, the use of uncertainty maps to support downstream analysis or subsequent machine learning tasks is widely used in various computer vision tasks(Skinner et al., 2021, Sajedi and Liang, 2021, Ning et al., 2021, Zhu et al., 2022). Further work is needed to explore the applications of uncertainty information to aid the process and analysis in qMRI.

In this work, we reported a deep learning-based framework for liver $T_{1\rho}$ mapping with uncertainty estimation to provide a direct estimation of the confidence level of $T_{1\rho}$ quantification and investigated the use of uncertainty information to improve the accuracy of $T_{1\rho}$ mapping of the liver using reduced number of $T_{1\rho}$-weighted images. The contributions are as follows:

1 We proposed formulating $T_{1\rho}$ mapping as a learning-based refinement problem that maps a coarse parametric map estimated from the conventional fitting approach to the final parametric map with simultaneous uncertainty estimation. We demonstrated this method improved the accuracy of $T_{1\rho}$ quantification compared to mapping $T_{1\rho}$-weighted images directly to a $T_{1\rho}$ map.

2 We proposed and demonstrated the uncertainty map can be used to assist in training an improved $T_{1\rho}$ mapping network (referred as improved network in the following context), which further improves the accuracy of $T_{1\rho}$ quantification of the liver.

3 We demonstrated that the uncertainty map can be utilized to discard unreliable areas in the liver parenchyma for ROI refinement to improve quantification accuracy.

## 2. METHODS

2.1 General framework

The mapping framework takes two $T_{1\rho}$ contrasts or two $T_{1\rho}$-weighted images as input and generates both the $T_{1\rho}$ map and the uncertainty map slice-by-slice. At each slice, the uncertainty map can be used to refine a user-defined ROI to improve the accuracy of the mean $T_{1\rho}$ value within the ROI by discarding pixels where $T_{1\rho}$ quantification is unreliable. The uncertainty map is also further processed and leveraged to train an improved network to improve $T_{1\rho}$ mapping accuracy. Figure 1 depicts the overall framework. The subsequent sections provide a detailed description of each component.

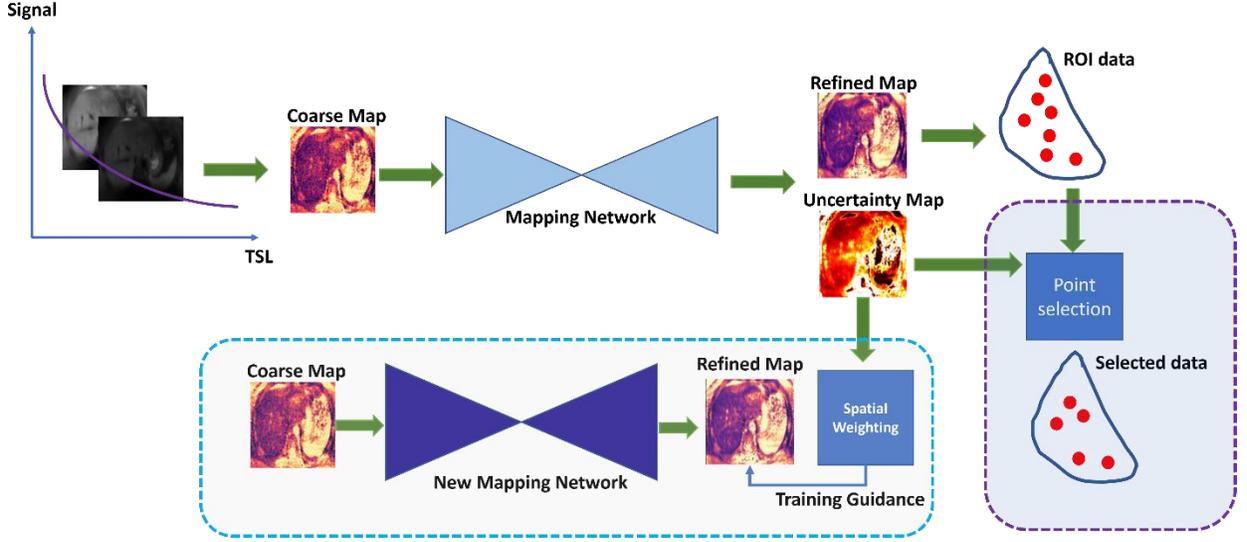

Figure 1: Overall pipeline of the uncertainty-aided learning-based $T_{1\rho}$ mapping and analysis framework. A coarse $T_{1\rho}$ map is first fitted using two $T_{1\rho}$ contrasts, and the coarse map is refined by an uncertainty-aware network. The acquired uncertainty map is further used for point selection in the post-hoc analysis and to improve the performance of an improved network.

2.2 Learning-based $T_{1\rho}$ map refinement

We formulate the $T_{1\rho}$ mapping task as a map refinement problem, which takes a coarse $T_{1\rho}$ map fitted by only two contrasts as input and produces a refined map. Given a mono-exponential model, the signal model of a $T_{1\rho}$ contrast image is shown in Equation 1:

$$S = S_0 \exp\left(-\frac{TSL}{T_{1\rho}}\right) \quad (1)$$

where $S_0$ is a scaling constant independent of TSL. The conventional methods typically apply least square fitting to multiple (usually more than 2) contrasts with different TSL to fit a $T_{1\rho}$ map. By using deep learning method, we aim to use only two $T_{1\rho}$ contrast images to estimate a reliable $T_{1\rho}$ map. Previous studies have demonstrated that utilizing only two contrasts to fit the $T_{1\rho}$ map via least square fitting method cannot produce accurate $T_{1\rho}$ measurement robustly(Huang et al., 2022a, Huang et al., 2022b), as shown in Figure 2. Nevertheless, compared to k-space data or $T_{1\rho}$ contrast images, the $T_{1\rho}$ map fitted from two $T_{1\rho}$-weighted images, which we term as coarse map in this work, contains more spatial information that the network can use as a prior to improve the optimization process to obtain a final $T_{1\rho}$ map. Concretely, the neural network takes the coarse map as the input and outputs a refined map. The values of the refined map are expected to be as close as possible to those maps fitted by the $N$ ($N > 2$) contrasts. The loss function can therefore be written as:

$$L = \frac{1}{ZP}\sum_{i=1}^{Z}\sum_{p=1}^{P}|\widehat{M}_{ip} - f_D\big(g(I_{xi}, I_{yi}); \theta\big)_p| \qquad (2)$$

where $\widehat{M}_i$ represents the well fitted map used as the target; $f_D(\cdot)$ is the deterministic mapping network with the trainable parameters denoted by $\theta$; $g(\cdot)$ is the least square fitting used to generate the coarse map; and $I_{xi}, I_{yi}$ are the two selected $T_{1\rho}$-weighted images. $p$ is the pixel index in each image. The pixel number in an image and the total number of slices in the training set is denoted as $P$ and $Z$ respectively. We omit the pixel index term in the following context to avoid text overlapping. We employ the $L_1$ norm as it is more robust to outliers in the target than $L_2$ norm. The flow, residual blood signal, and MRI system imperfections such as $B_1$ RF and $B_0$ field inhomogeneity sometimes can lead to exceptionally low or high $T_{1\rho}$ estimations beyond the range of physiological $T_{1\rho}$ values. We empirically found that directly using $T_{1\rho}$ maps including these values make it challenging for the model to converge. To address this issue, we set the thresholds of the estimated $T_{1\rho}$ values of the liver within a cut-off range of 25ms to 65ms based on the liver $T_{1\rho}$ values previously reported(Wáng et al., 2018, Hou et al., 2023).

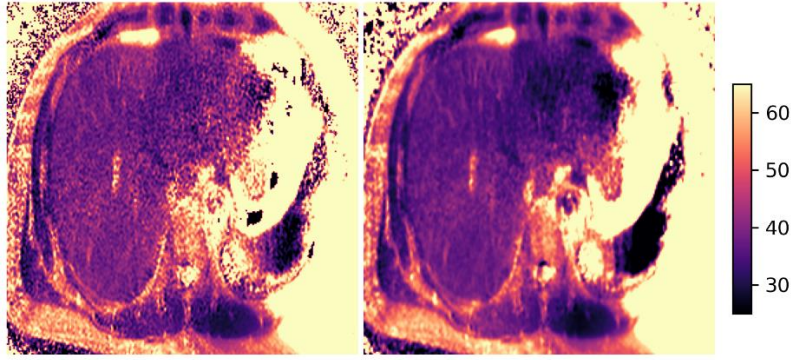

Figure 2: Comparison example of the coarse map and the map fitted by four contrasts (target). Left: The coarse map. Right: The map fitted by four contrasts. The unit of the color bar is in millisecond.

2.3 Uncertainty estimation

Let us denote the input of the probabilistic $T_{1\rho}$ mapping network $f_P(\cdot)$ as $X_i$, with the output of the network being $\{M_i, J_i \cdots K_i\} = f_P(X_i; \theta)$. $\{M_i, J_i \cdots K_i\}$ represents the parameters of the output distribution $\widehat{M}_i \sim P(\widehat{M}; \{M_i, J_i \cdots K_i\})$, where $\widehat{M}$ represents the target and $i$ denotes the index of a sample in the dataset. Note that $M_i$ is the predicted map of the neural network. To estimate the parameter of the output distribution, we need to minimize the negative likelihood function of the dataset. The likelihood function can be expressed as follows:

$$\Gamma = \prod_{i=1}^{Z} P(\widehat{M}; \{M_i, J_i \cdots K_i\}) \tag{3}$$

By selecting an appropriate distribution model, we can estimate the uncertainty based on the output distribution parameters of the network. We opt for the Laplace distribution to model the distribution as it contains an $L_1$ norm term that is consistent with that in Equation (2). Note that the symmetrical attribute of Laplace distribution makes it suitable as an alternative when the $L_2$ norm of the Gaussian distribution is affected by the outliers in the target. It is also common to adopt the Laplace distribution to model the uncertainty in those conventional computer vision or image processing tasks(Su et al., 2022, Dong et al., 2023). By plugging the Laplace distribution into Equation (3) and performing the negative logarithmic operation, we have:

$$-\log \prod_{i=1}^{Z} \frac{1}{2\sigma_i} \exp\left(-\frac{|\widehat{M}_i - M_i|}{\sigma_i}\right) \propto \sum_{i=1}^{Z} \left(\frac{|\widehat{M}_i - M_i|}{\sigma_i} + \log(\sigma_i)\right) \quad (4)$$

where $\sigma_i$ represents the scale parameter of the Laplace distribution, which indicates the level of uncertainty in the predicted value. In the Laplace distribution, $\sqrt{2}\sigma_i$ corresponds to the standard deviation of the predicted distribution, which we use as the final uncertainty map. Note that $\sigma_i$ is a learnable parameter of the $T_{1\rho}$ mapping network. To ensure numerical stability, the network is trained to directly learn the logarithmic term in the loss function(Kendall and Gal, 2017) and the loss function becomes:

$$L = \sum_{i=1}^{Z} \left(\exp(-S_i)|\widehat{M}_i - M_i| + S_i\right) \quad (5)$$

By utilizing the loss function in Equation (5) during the optimization of the mapping network, the network can be used to predict the $T_{1\rho}$ value and the corresponding uncertainty once trained. It should be noted that this loss function can spatially weigh different areas during the learning process. Areas with erroneous values are typically associated with noisy labels that can be difficult to learn, which results in increased uncertainty values at these areas to attenuate their contributions when minimizing the loss function. The additional added term $S_i$ can regularize the model to avoid learning large uncertainty values at all areas. The resulted model is a probabilistic version of the $T_{1\rho}$ mapping network introduced in section 2.2.

2.4 Uncertainty aided improved network

The uncertainty map obtained using the aforementioned model can be utilized to train an improved network to further improve $T_{1\rho}$ quantification of the liver parenchyma. To achieve this, we first convert the uncertainty map into a spatial weighting map, which is then utilized to guide the learning process. Although it may seem intuitive to use the obtained uncertainty map directly as the spatial weighting map, it is important to note that the areas with cut off values typically have low uncertainty, which results in large weights at these areas if we use uncertainty for weighting directly. Since the pixels with cut-off $T_{1\rho}$ values are trivial for the network to learn and not of interests of the user, we do not want the network to prioritize them during the learning process. To address this issue, we assign the largest uncertainty value in the uncertainty map to those areas which satisfy the following conditions: 1) having extremely low uncertainty (below a certain threshold) in the uncertainty map and 2) having the cut off values in the target $T_{1\rho}$ map. The uncertainty values at the other regions are unchanged. By doing so, the liver parenchyma area has relatively low uncertainty values in the uncertainty map and can be prioritized during training. Mathematically, the resulting spatial weighting is applied to the $S_i$ map in Eq.(5) to create a spatial map $\bar{S}_i$, which is plugged into the following loss function to train the improved model:

$$L = \sum_{i=1}^{Z} \exp(-\bar{S}_i) |\widehat{M}_i - M_i| \qquad (6)$$

2.5 Uncertainty-aided ROI refinement

We also use the uncertainty map to refine the ROIs for $T_{1\rho}$ measurement of the liver parenchyma. In our framework, ROIs are firstly obtained on the anatomical image, after which the algorithm assists in refining the ROIs by filtering out pixels with low confidence levels indicated by their uncertainty values. The uncertainty threshold is computed in the validation set and is determined by taking the mean uncertainty value of all pixels in all ROIs and added with the corresponding standard deviation of the uncertainty values in the ROIs, denoted as $\mu_{uncer} + std_{uncer}$. During inference, those pixels with uncertainty values larger than the threshold are discarded.

2.6 Experiment design

Our experiment includes two parts. In the first part, we compared the performance of different methods in $T_{1\rho}$ predictions in the liver of patients at different stages of non-alcoholic fatter liver disease. Specifically, we evaluate the following methods:

Coarse fitting: This method involves directly fitting the $T_{1\rho}$ map using two $T_{1\rho}$ contrasts via a non-linear least square fitting method without any post-processing.

BM3D: The BM3D(Dabov et al., 2007) method is a classic non-learning-based image denoising technique. Since the map refinement task is similar to the image restoration task, we apply BM3D to the coarse map and use the resulting denoised map as the final output.

Learning based refinement (LBR): This is our proposed method that utilizes the deterministic loss function in Equation (2).

Learning based refinement with uncertainty (LBRU): This is the proposed probabilistic mapping method. We train the probabilistic $T_{1\rho}$ mapping network with the loss function in Equation (4).

Learning based mapping from $T_{1\rho}$ contrast images (LBMC): To demonstrate the advantages of the proposed mapping refinement approach which uses a coarse $T_{1\rho}$ map as an input to the network, we also trained a mapping network that takes two $T_{1\rho}$-weighted images as the input and outputs a $T_{1\rho}$ map. The architecture of this model is similar to that of the LBR, except that the input is a stacked two-channel tensor.

Learning based uncertainty driven refinement (LBUDR): To examine the effectiveness of using uncertainty map to further improve the performance of the model, we also test the uncertainty-aided method for training the improved network as described in section 2.4. For the improved network, we maintain the same network architecture as that of the LBR.

In the second part of our experiment, we provided a systematic analysis of the estimated uncertainty, examined the effectiveness of ROI refinement using acquired uncertainty map. Note that the ROI refinement is done on the trained model of the LBRU, as the uncertainty map reflects the confidence level of the predicted $T_{1\rho}$ map of this model.

2.7 Evaluations

We used the following metrics in this work:

ROI Mean Relative Error (RMRE):

The performance evaluation was carried out in the human drawn ROI and the ROI drawing follows the principle in previous study(Huang et al., 2022b) by avoiding large blood vessels and bile ducts. The RMRE is defined as the mean relative error of the pixel value of the $T_{1\rho}$ map in the ROI:

$$RMRE = \frac{1}{Z}\sum_{z=1}^{Z}\frac{1}{N}\sum_{n=1}^{N}\frac{|T_{1\rho n}-\widehat{T_{1\rho n}}|}{\widehat{T_{1\rho n}}} \times 100\% \quad (T_{1\rho n},\widehat{T_{1\rho n}})\in ROI_z \quad (7)$$

where $Z$ and $N$ are the number of slice and the number of pixels within the ROI respectively. $T_{1\rho n}$ and $\widehat{T_{1\rho n}}$ are the predicted value and the target value of a pixel respectively.

Uncertainty Calibration Error (UCE):

The UCE measures how well the uncertainty reflects the level of the absolute error(Upadhyay et al., 2022). Denoting the mean value in the ROI of each slice as the slice value, we compute the slice value of the uncertainty map and the absolute error map in the ROI of every slice and use these two to compute the UCE. We bin the slices into a histogram according to the mean uncertainty in the ROIs, and we compute the UCE in the following way(Laves et al., 2020):

$$UCE = \sum_{m=1}^{M}|B_m|\frac{|err(B_m)-uncer(B_m)|}{N} \quad (8)$$

$$err(B_m) = \frac{1}{|B_m|}\sum_{i \in B_m} |p_i - \hat{p}_i| \qquad (9)$$

$$uncer(B_m) = \frac{1}{|B_m|}\sum_{i \in B_m} u_i \qquad (10)$$

$B_m$ stands for the $m^{th}$ bin and $|B_m|$ is the number of slices in that bin. $p_i$ and $\hat{p}_i$ represent the mean $T_{1\rho}$ value in the ROI of the predicted map and the target map. Similarly, $u_i$ is the mean uncertainty value in the ROI of the predicted map. A lower UCE indicates better estimation of uncertainty, as it reflects the level of the error.

We also used the reliability diagram(Laves et al., 2019) to evaluate the discrepancy of the estimated uncertainty and the actual absolute error qualitatively. It is a plot of the $err(B_m)$ according to $uncer(B_m)$ for each bin. If the uncertainty is modelled properly, the plotted curve should be close to the ideal curve of y = x.

In addition, we conducted statistical analyses for hypothesis testing. A pair t-test was used to test the null hypothesis that there was no significant difference between the slice value of the absolute error from the predicted map after point selection according to the uncertainty threshold and that before point selection.

We also used the sparsification plot (Wang et al., 2022) to mimic the removal of the unconfident pixels in the ROIs according to the ranking of the uncertainty values qualitatively. We sorted all pixels of the predicted $T_{1\rho}$ map in the ROIs in descending order of uncertainty and iteratively remove a subset of pixels (top 5% in this work). We then computed the mean RMRE of the remaining pixels to plot a curve. An ideal curve (oracle curve) was obtained by ranking the pixels in a descending trend according to the relative error, while a random curve was modeled by removing pixels randomly and is expected to be a flat curve. If the uncertainty map can effectively be used to remove pixels with high error, the curve should have a descending trend similar to the ideal curve.

2.8 Data acquisition and datasets

The scan was approved by the institutional review board and was conducted throughout the year 2019. The datasets of 51 patients (27 males, 24 females) at different non-alcoholic fatty liver diseases stages was retrospectively used in this study. The demographic of the dataset is shown in Table 1. All the data were collected on a 3.0 T MRI scanner (Philips Achieva, Philips Healthcare, Best, Netherland). A body coil was used as the RF transmitter, and with a 32-channel cardiac receiver coil(In vivo Corp, Gainesville, USA). The datasets were collected using breath-hold magnetization prepared turbo spin echo acquisition with suppression of blood signal(Chen et al., 2016). Pencil-beam shimming was applied on the liver parenchyma to reduce $B_0$ field inhomogeneity around the liver parenchyma. The vendor-provided RF shimming was applied to reduce $B_1$ RF inhomogeneity. The $T_{1\rho}$ contrast images were collected using the time of spin-lock (TSL) of 0, 10, 30, 50 ms, respectively. Three slices were collected from each subject. The scan parameters include: resolution = $1.5 \times 1.5\ mm^2$, slice thickness = 7mm, time of repetition = 2000 ms, echo time = 10 ms, and Frequency of spin-lock = 400 Hz. Fat signal was suppressed using Spectral Attenuated Inversion Recovery (SPAIR). The data was split into three groups for a three-fold validation. In each group, the data of 17 patients are used for testing and the rest of the data are used for model development. Among the rest of the data in each group for model development, the data of 30 patients are used for training and four for validation. The results are reported as the average of the results from the three groups.

|  | Gender statistics | |
|---|---|---|
|  | Number | Mean age |
| Female | 24 | 60.29±9.88 years |
| Male | 27 | 54.79±8.95 years |
|  | Liver Fibrosis stage statistics | |
|  | Number | Mean age |
| F0 | 25 | 57.60 ± 9.75 years |
| F1 | 13 | 56.77 ± 7.65 years |
| F2 | 14 | 57.36 ± 11.45 years |
| Total | 51 | 57.33 ± 9.79 years |

Table 1: Dataset demographics

2.9 Network architecture

For the deterministic $T_{1\rho}$ mapping network, we utilized a UNet-like structure. The network is designed for a single-channel map refinement task with both the input and output layers consisting of one channel. Similarly, we employed the same architecture for the probabilistic mapping network, but with a modified output layer consisting of two channels, one for the uncertainty map and the other for the $T_{1\rho}$ mapping. We include the architecture details in the supplementary material.

2.10 Implementation details

We selected the contrasts with TSL = 0 and 50 ms to calculate the coarse $T_{1\rho}$ map by using least square fitting, as previous work shown such choice of TSLs can produce a parametric map relatively close to the target map(Huang et al., 2022b). All images were resized to matrix size 256 × 256. Data augmentation was performed by random rotation ranging from -7 to 7 degrees and random translation ranging from -10 to 10 pixels in four vertical directions. Note the target maps after augmentation are fitted by those augmented $T_{1\rho}$ contrasts. The learning rate was set to 1e-3. The batch size was 4. The optimizer used was ADAM(Kingma and Ba, 2014). The bin number of UCE was set as 7. The experiments were carried out on a Nvidia RTX 3080 Ti GPU with AMD Ryzen 9 5900X 12-Core CPU (3.70 GHz), and the experiment framework was Pytorch 1.9(Paszke et al., 2019). The statistical analysis was carried out using the Python Scipy package(Virtanen et al., 2020).

## 3. RESULTS

Table 2 presents the results of different mapping methods compared in this study. All our proposed learning-based refinement methods achieved errors of less than 3%, whereas the other methods yielded errors of over 4%. The proposed deterministic method (LBR) achieved slightly better results than the proposed probabilistic method (LBRU) (2.71% vs. 2.80%). The reason behind this phenomenon will be discussed in the following sections. By using the uncertainty aided method, the proposed LBUDR achieved the best performance (2.60%).

| Method | RMRE (%) |
|---|---|
| Coarse fitting | 4.88±2.12 |
| BM3D | 4.82±2.45 |
| LBR | 2.71±1.32 |
| LBRU | 2.80±1.21 |
| LBMC | 4.28±2.68 |
| LBUDR | **2.62±1.15** |

Table 2: The RMRE of different methods for fitting the $T_{1\rho}$ map.

Figure 3 shows Bland-Altman plots of the $T_{1\rho}$ pixel values predicted in the ROI from all three groups of the cross-validation. All learning-based methods have a mean error centered near zero and a relatively small standard deviation of the error. The other three methods show a larger bias as well as a larger error standard deviation.

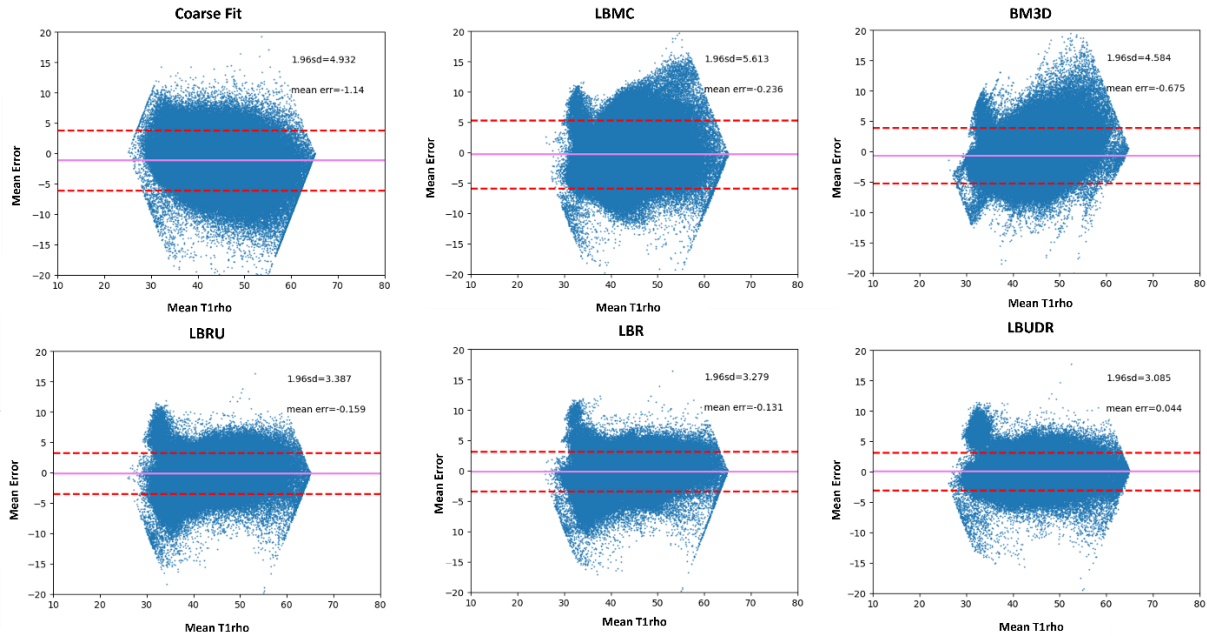

Figure 3: The Bland-Altman plot of the pixel-wise $T_{1\rho}$ value of different methods for fitting the $T_{1\rho}$ map. The unit is in millisecond.

Figure 4 shows typical visualization examples of the predicted $T_{1\rho}$ map of different methods, which visually verifies that the learning-based refinement methods provide better prediction performance in the parenchyma area.

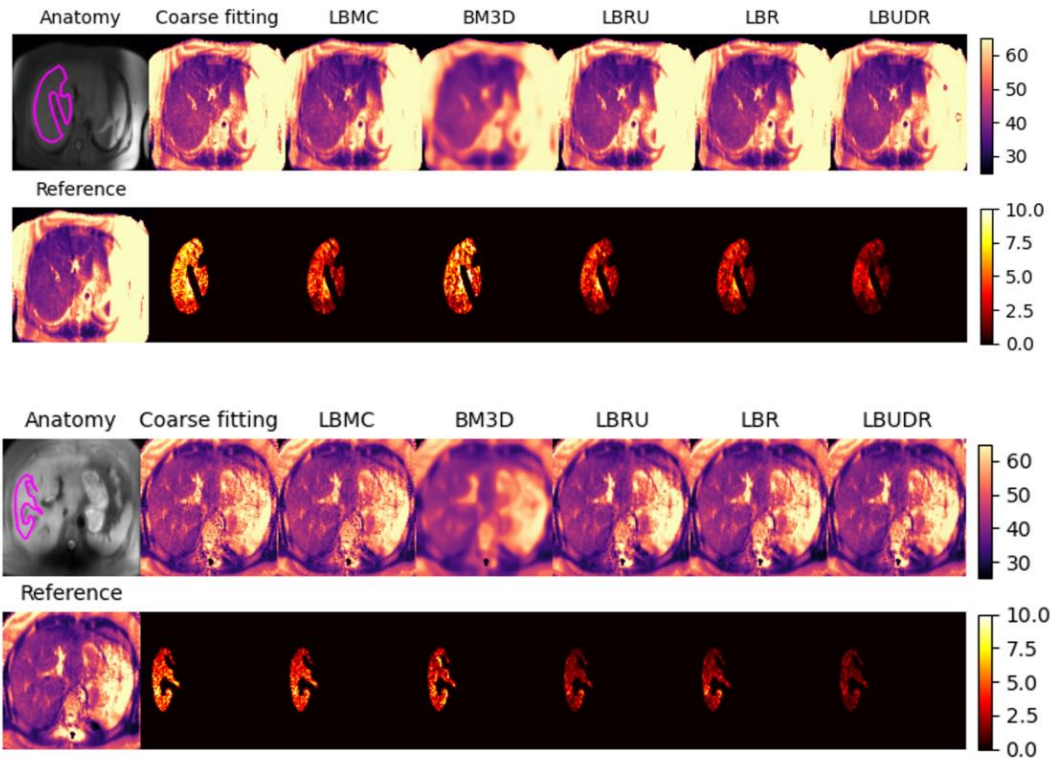

Figure 4: Two visualization examples of the predicted maps and the corresponding error maps in the ROI. Starting from the second column from the left, the images in the first row are the predicted maps of different methods (unit in ms) and those in the second row are the corresponding relative error maps (unit in percentage).

Figure 5 shows a scatter plot of the predicted pixel values and the target pixel values of all three groups of the cross-validation of the proposed LBRU, indicating a reasonable agreement between the prediction and the target values, even though the prediction performance was slightly inferior to the proposed deterministic version, the LBR.

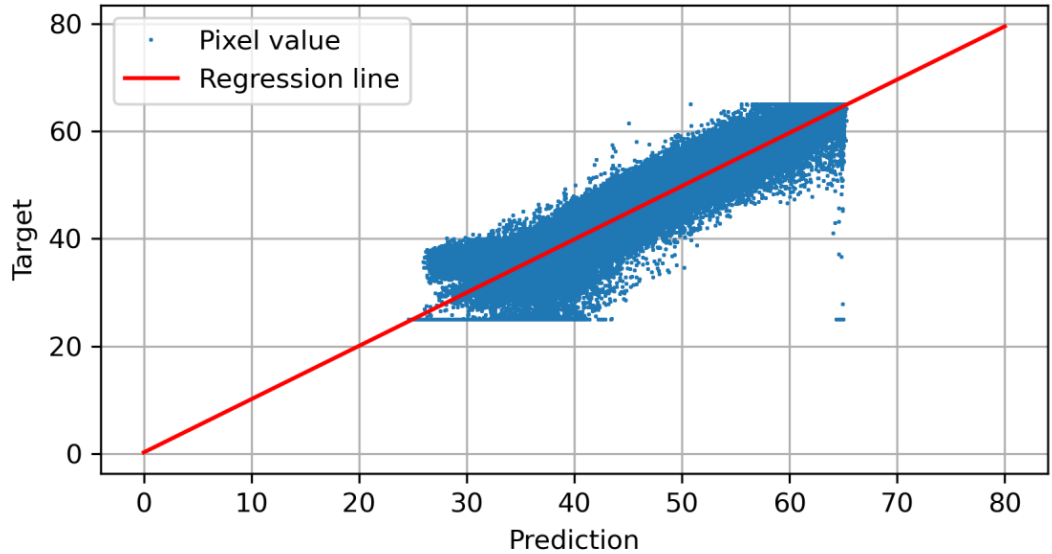

Figure 5: Linear regression line of all the predicted pixel value of LBRU and the target. The fitted result is $y = 0.9905x + 0.2408$. The Pearson Correlation Coefficient is 0.95. The standard error of the slope is 3.46e-4. The unit is in milliseconds.

Figure 6 shows the reliability diagram of the proposed LBRU approach. The reliability curve is close to the ideal curve with an overall slight overestimation, indicating that the uncertainty quantification reflects the level of the actual error. The UCE of the estimated uncertainty is 0.09 ms, which also demonstrates that the estimated uncertainty has a decent indication of the error level.

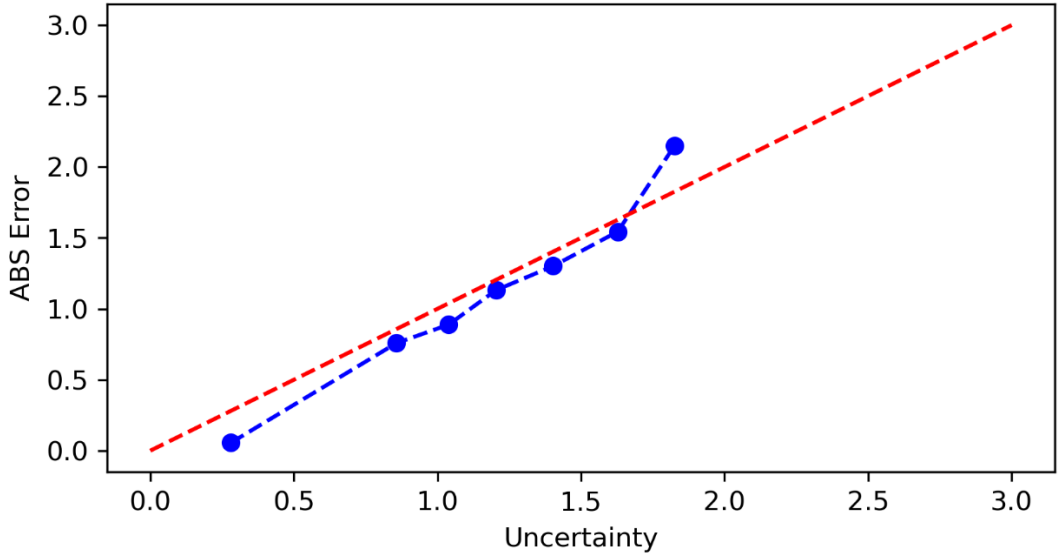

Figure 6: Reliability diagram of the uncertainty estimation of LBRU. The unit is in millisecond. The blue curve is plotted according to the value in the binned histogram while the red curve with dashed line is the ideal curve (y=x).

Figure 7 presents an illustrative example of the uncertainty map estimated by the LBRU. The map displays a reasonable spatial distribution of the uncertainty level, as compared to the actual absolute error map. It can also be seen the liver parenchyma in the right lobe of the liver had a relatively lower uncertainty level of $T_{1\rho}$ quantification than other tissues like the blood vessels, bile ducts and those tissues outside the shimming areas where there was exacerbated field inhomogeneity after applying the shimming and thus unreliable $T_{1\rho}$ measurement. Also, those areas with cutoff values of $T_{1\rho}$ had low uncertainty values, which may distract the model from learning the representation of the liver parenchyma if they are not properly handled during the optimization.

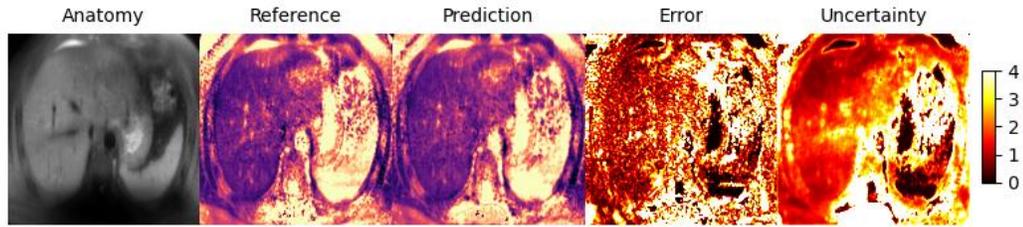

Figure 7: A typical example of the estimated uncertainty map. The unit of both the absolute error map and the uncertainty map are in milliseconds.

Figure 8 shows typical examples of ROI refinement based on uncertainty. The figure indicates that areas with relatively high absolute error are often associated with relatively high uncertainty values. Consequently, the uncertainty map is effective to guide the selection of points within the ROI for ROI refinement to improve reliability of $T_{1\rho}$ measurement of the liver.

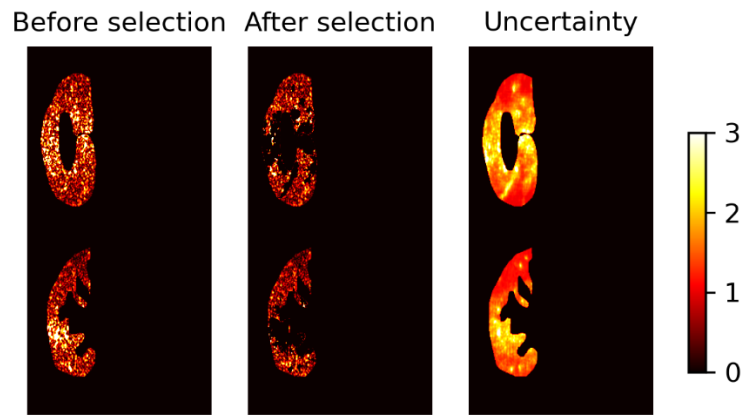

Figure 8: Typical examples of point selection in the ROIs according to uncertainty. From left to right: Absolute error map before point selection, absolute error maps after point selection according to the uncertainty maps, uncertainty maps. The unit of the color bar is in millisecond.

The $p$-value of the hypothesis testing for the null hypothesis is 1.15e-16 < 0.05 (1.06 ±0.68 ms vs 1.14 ±0.76 ms), which rejects the null hypothesis and suggests that the slice value of the absolute error from the predicted map, after point selection according to the uncertainty threshold, is significantly lower than that before point selection.

Figure 9 depicts the sparsification plot of the estimated uncertainty of the LBRU model. The sparsification curve demonstrates a descending trend similar to the oracle curve, and lies below the random curve, indicating a reasonable agreement between the ranking of the uncertainty and the relative error.

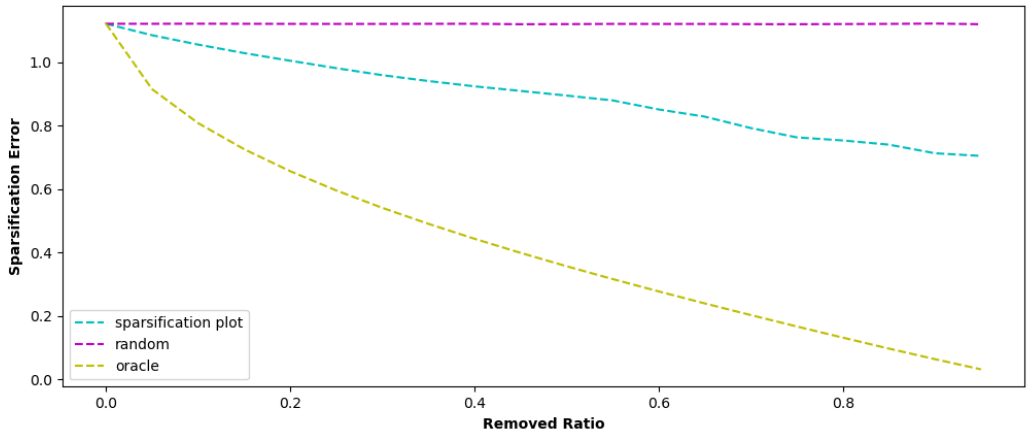

Figure 9: Sparsification plot of removing pixels in the ROI according to the estimated uncertainty of LBRU

## 4. DISCUSSION

In this study, we addressed $T_{1\rho}$ quantification by formulating it as a learning-based map refinement problem. We achieved an error level of less than 3% in predicting $T_{1\rho}$ values from only two $T_{1\rho}$-weighted images in the liver parenchyma, which demonstrates potential of saving scan time while preserving quantification accuracy. For our liver $T_{1\rho}$ imaging protocol, reducing the number of $T_{1\rho}$-weighted images from 4 to 2 will result in a scan time reduction from 10 seconds to 6 seconds, given a typical TR of 2 seconds. This reduction in scan time is especially beneficial for breath-hold imaging, particularly for individuals who have difficulties holding their breath. This time-saving advantage can also be leveraged to enhance the volume coverage of liver imaging. A recent study(Qian et al., 2023) demonstrated that dynamic $T_{1\rho}$ imaging successfully detects signal changes associated with glucose level variations in the liver following oral glucose ingestion at clinical dosage. Our learning-based method can be used to increase temporal resolution and volume coverage of dynamic glucose enhanced $T_{1\rho}$ imaging of the liver. The proposed method can be extended to 3D $T_{1\rho}$ imaging of other anatomical regions such as the knee and brain. The impact of scan time reduction offered by our learning-based approach can be more significant for 3D $T_{1\rho}$ imaging, as each $T_{1\rho}$-weighted volume can take several minutes to acquire.

Our work demonstrates that a $T_{1\rho}$ mapping network that uses $T_{1\rho}$-weighted images (LBMC) as input produces inferior results compared to the proposed approach (LBR). One possible explanation is that the domain gap between the $T_{1\rho}$-weighted images and $T_{1\rho}$ map is relatively large, and the network needs to implicitly learn the exponential decay principle for mapping the $T_{1\rho}$. It is possible that the network overfits on the information which is irrelevant to the mapping task in the image domain. In contrast, our learning-based refinement method provides spatial prior knowledge of the parametric map to the model, resulting in a small domain gap between the input and the output. Overall, our approach represents a promising solution for accelerating quantitative MRI by using reduced number of contrast images.

Our study demonstrates that incorporating uncertainty estimation in the $T_{1\rho}$ mapping network can improve the reliability of predicted values. This is particularly important for quantitative $T_{1\rho}$ imaging since the ground truth is not available during testing. The uncertainty map can also provide a guide for human-computer interactions. Our experiments demonstrated that the estimated uncertainty map from the network has potential post-hoc applications, such as ROI refinement and providing spatial weighting to train models to further improve $T_{1\rho}$ mapping.

We observed the proposed LBRU model which includes the uncertainty estimation during the training leads to a marginal degradation in $T_{1\rho}$ prediction in the liver parenchyma compared to the proposed deterministic version. One possible reason is that the areas with cut-off values in the targets are "very certain" during training and have a higher weighting, which can distract the model from the parenchyma during training. The improved performance of the newly trained model based on the processed spatial weighting map (LBUDR) further supports this explanation, as the cut-off areas are assigned the largest uncertainty value during the new model training. Despite the slight drop in mapping performance of LBRU, our experimental results show that the predicted results are still in decent agreement with the target, while simultaneously acquiring uncertainty.

There are limitations in this work. We studied liver disease of early-stage liver fibrosis, which has a relatively homogeneous structure and relatively insignificant alterations of relaxation rates. More severe liver diseases may have increased change of $T_{1\rho}$ values. Anatomies like brain and knee have more complicated structures. Further investigation is needed to extend the proposed methods for applications in other liver diseases and other anatomies. Another limitation is that our method still relies on supervised learning to obtain the confidence level of the quantification results. Supervised learning requires high-quality labeled target data which can be difficult to obtain. Future work is needed to obtain unit consistent uncertainty of tissue parameter quantifications based on self-supervised learning, such as the approach recently reported in computer vision tasks(Dikov and van Vugt, 2022).

## 5. CONCLUSION

Our proposed learning-based method for refining $T_{1\rho}$ maps can generate both refined parametric maps and corresponding estimation of uncertainty levels using only two $T_{1\rho}$-weighted images. Our method produces $T_{1\rho}$ values that are well aligned with target maps obtained from four $T_{1\rho}$-weighted images, and the uncertainty maps reflect the level of quantification error. We demonstrated the estimated uncertainty map can be used for ROI refinement in the liver and as spatial weighting to further improve the accuracy of $T_{1\rho}$ quantification of the liver.

**Acknowledgement:**

This study was supported by a grant from the Research Grants Council of the Hong Kong SAR (Project GRF 14201721), a grant from the Innovation and Technology Commission of the Hong Kong SAR (Project No.MRP/046/20x). The author would also like to thank Dr Yurui Qian, Dr Jian Hou and Dr Baiyan Jiang for their assistance in data acquisition and useful daily discussions.

**Conflict of interests:**

None of the authors have conflict of interests.

Diagnostic potential of T1ρ and T2 relaxations in assessing the severity of liver fibrosis and necro-inflammation. *Magnetic Resonance Imaging,* 87**,** 104-112.

UPADHYAY, U., KARTHIK, S., CHEN, Y., MANCINI, M. & AKATA, Z. BayesCap: Bayesian Identity Cap for Calibrated Uncertainty in Frozen Neural Networks.  Computer Vision–ECCV 2022: 17th European Conference, Tel Aviv, Israel, October 23–27, 2022, Proceedings, Part XII, 2022. Springer, 299-317.

VIRTANEN, P., GOMMERS, R., OLIPHANT, T. E., HABERLAND, M., REDDY, T., COURNAPEAU, D., BUROVSKI, E., PETERSON, P., WECKESSER, W. & BRIGHT, J. 2020. SciPy 1.0: fundamental algorithms for scientific computing in Python. *Nature methods,* 17**,** 261-272.

WANG, C., WANG, X., ZHANG, J., ZHANG, L., BAI, X., NING, X., ZHOU, J. & HANCOCK, E. 2022. Uncertainty estimation for stereo matching based on evidential deep learning. *Pattern Recognition,* 124**,** 108498.

WÁNG, Y. X. J., DENG, M., LIN, J., KWOK, A. W., LIU, E. K. & CHEN, W. 2018. Age-and Gender-Associated Liver Physiological T1rho Dynamics Demonstrated with a Clinically Applicable Single-Breathhold Acquisition. *SLAS technology,* 23**,** 179-187.

ZHU, Y., DONG, W., LI, L., WU, J., LI, X. & SHI, G. Robust depth completion with uncertainty-driven loss functions.  Proceedings of the AAAI Conference on Artificial Intelligence, 2022. 3626-3634.
29